\documentclass[aip,jcp,reprint,noshowkeys,superscriptaddress]{revtex4-2}
\usepackage{graphicx,dcolumn,bm,xcolor,microtype,multirow,amscd,amsmath,amssymb,amsfonts,physics,longtable,wrapfig,bbold,siunitx,xspace}
\usepackage[version=4]{mhchem}

\usepackage[utf8]{inputenc}
\usepackage[T1]{fontenc}

\usepackage{hyperref}
\hypersetup{
    colorlinks,
    linkcolor={red!50!black},
    citecolor={red!70!black},
    urlcolor={red!80!black}
}

\usepackage{listings}
\definecolor{codegreen}{rgb}{0.58,0.4,0.2}
\definecolor{codegray}{rgb}{0.5,0.5,0.5}
\definecolor{codepurple}{rgb}{0.25,0.35,0.55}
\definecolor{codeblue}{rgb}{0.30,0.60,0.8}
\definecolor{backcolour}{rgb}{0.98,0.98,0.98}
\definecolor{mygray}{rgb}{0.5,0.5,0.5}

\definecolor{sqred}{rgb}{0.85,0.1,0.1}
\definecolor{sqgreen}{rgb}{0.25,0.65,0.15}
\definecolor{sqorange}{rgb}{0.90,0.50,0.15}
\definecolor{sqblue}{rgb}{0.10,0.3,0.60}

\lstdefinestyle{mystyle}{
    backgroundcolor=\color{backcolour},
    commentstyle=\color{codegreen},
    keywordstyle=\color{codeblue},
    numberstyle=\tiny\color{codegray},
    stringstyle=\color{codepurple},
    basicstyle=\ttfamily\footnotesize,
    breakatwhitespace=false,
    breaklines=true,
    captionpos=b,
    keepspaces=true,
    numbers=left,
    numbersep=5pt,
    numberstyle=\ttfamily\tiny\color{mygray},
    showspaces=false,
    showstringspaces=false,
    showtabs=false,
    tabsize=2
  }

  \usepackage[]{notes2bib}

  \newcolumntype{d}{D{.}{.}{-1}}

\newcommand{\ie}{\textit{i.e.}\xspace}

\newcommand{\hH}{\Hat{H}} 
\newcommand{\hT}{\Hat{T}} 
\newcommand{\hV}{\Hat{V}} 
 
\newcommand{\hW}{\Hat{W}}

\newcommand{\hn}{\Hat{n}}
\newcommand{\cre}[1]{a_{#1}^\dagger}
\newcommand{\ani}[1]{a_{#1}} 

\newcommand{\br}{\boldsymbol{r}}

\newcommand{\up}{\uparrow} 
\newcommand{\dw}{\downarrow} 
\newcommand{\Dv}{\Delta v} 
\newcommand{\Dn}{\Delta n} 
\DeclareMathOperator*{\stat}{stat}
\newcommand{\Ts}{T_\text{s}} 
 
\newcommand{\Tc}{T_\text{c}} 
 
\newcommand{\Eh}{E_\text{H}} 
\newcommand{\Ex}{E_\text{x}} 
\newcommand{\Ec}{E_\text{c}} 
\newcommand{\Exc}{E_\text{xc}} 
 
\newcommand{\EKS}{E_\text{KS}} 
 
\newcommand{\Wc}{W_\text{c}} 
\newcommand{\Ehxc}{E_\text{Hxc}} 
\newcommand{\vhxc}{v_\text{Hxc}}

\newcommand{\vh}{v_\text{H}}
\newcommand{\vx}{v_\text{x}}
\newcommand{\vc}{v_\text{c}}
\newcommand{\vs}{v_\text{s}}

\newcommand{\cmel}[3]{\mel*{#1}{#2}{#3}_\text{c}}
\newcommand{\cbraket}[2]{\braket*{#1}{#2}_\text{c}}

\newcommand{\ii}{\mathrm{i}}

\newcommand{\LCPQ}{Laboratoire de Chimie et Physique Quantiques (UMR 5626), Universit\'e de Toulouse, CNRS, UPS, France}
\newcommand{\ICP}{Institut de Chimie Physique (UMR 8000), Universit\'e Paris-Saclay, CNRS}
\begin{document}	

\title{Excited-State-Specific Kohn-Sham Formalism for the Asymmetric Hubbard Dimer}

\author{Pierre-Fran\c{c}ois \surname{Loos}}
	\email{loos@irsamc.ups-tlse.fr}
	\affiliation{\LCPQ}

\author{Sara \surname{Giarrusso}}
	\email{sara.giarrusso@universite-paris-saclay.fr}
	\affiliation{\ICP}

\begin{abstract}
Building on our recent study [\href{https://doi.org/10.1021/acs.jpclett.3c02052}{\textit{J.~Phys.~Chem.~Lett.}~\textbf{14}, 8780 (2023)}], we explore the generalization of the ground-state Kohn-Sham (KS) formalism of density-functional theory (DFT) to the (singlet) excited states of the asymmetric Hubbard dimer at half-filling. 
While we found that the KS-DFT framework can be straightforwardly generalized to the highest-lying doubly-excited state, the treatment of the first excited state presents significant challenges. 
Specifically, using a density-fixed adiabatic connection, we show that the density of the first excited state lacks non-interacting $v$-representability. 
However, by employing an analytic continuation of the adiabatic path, we demonstrate that the density of the first excited state can be generated by a complex-valued external potential in the non-interacting case.
More practically, by performing state-specific KS calculations with exact and approximate correlation functionals --- each state possessing a distinct correlation functional --- we observe that spurious stationary solutions of the KS equations may arise due to the approximate nature of the functional.
\end{abstract}

\maketitle
%%%%%%%%%%%%%%%%%%%%%%%%
\section{Introduction}
\label{sec:intro}
%%%%%%%%%%%%%%%%%%%%%%%%

The Kohn-Sham (KS) formalism \cite{KohSha-PRA-65} of density-functional theory (DFT) \cite{HohKoh-PR-64} is a cornerstone of electronic structure theory, celebrated as the method of choice across chemistry, physics, and materials science for its practicality and versatility. \cite{TeaHelSav-PCCP-22} 
At its heart lies the KS model system, a hypothetical construct of non-interacting electrons designed to reproduce the electron density of the corresponding interacting system. \cite{KohSha-PRA-65}
To recreate the electron density of the physical system, the KS system evolves in a local multiplicative one-body effective potential referred to as the KS potential, defined up to an arbitrary additive constant.
While the Hohenberg-Kohn theorem \cite{HohKoh-PR-64} does not ensure the existence of this potential, it does establish its uniqueness for ground-state systems.

The central issue of the existence of such a potential is known as the $v$-representability problem. \cite{Koh-PRL-83,Lam-PRA-10}
A density is $v$-representable if it arises from a $N$-electron system with an external potential $v$. 
Densities are further classified as interacting or non-interacting $v$-representable, depending on whether they arise from an interacting or a non-interacting system, respectively.
For the KS formalism to be applicable, the interacting $v$-representable density of the physical electron system must also be non-interacting $v$-representable --- a necessary condition.
Thus, the focus is on densities that are both interacting and non-interacting $v$-representable.

There is clear evidence that certain physical ground states lack a corresponding pure-state KS system, \cite{SchGriBae-TCA-98, TruErhGor-PRA-24} pointing out fundamental limitations of the KS formalism.\cite{Lee-AQC-03} 
The existence of a pure-state KS system for excited states remains even more uncertain. \cite{Gor-PRA-96,Gor-PRA-99,Gar-ARMA-22}
The Hohenberg-Kohn theorem guarantees a bijective mapping between the external potential $v$ and the ground-state density $\rho$  \cite{HohKoh-PR-64,Gar-MPAG-18,PenTelCsiRugLae-ACSPCAu-23} but no analogous theorem exists for excited states. \cite{GauBur-PRL-04,SamHarHol-CPL-06,SamHar-JPB-06,Gar-CMP-21}

Despite this, orbital-optimized DFT (OO-DFT) \cite{Gor-PRA-96,Nag-IJQC-98,LevNag-PRL-99,Gor-PRA-99,ZhaBur-PRA-04,AyeLev-PRA-09,AyeLevNag-PRA-12,AyeLevNag-JCP-15,AyeLevNag-TCA-18,Gar-ARMA-22} has gained traction for excited-state calculations as it consistently delivers accurate excitation energies, particularly for double excitations and charge-transfer states. \cite{PerLev-PRB-85,FilSha-CPL-99,GilBesGil-JPCA-08,KowTsuTak-JCP-13,LevIvaJon-JCTC-20,LevIvaJon-FD-20,HaiHea-JCTC-20,HaiHea-JPCL-21,BarGilGil-JCTC-18a,BarGilGil-JCTC-18b,CarHer-JCTC-20,SchLevJon-JCTC-23,SelSigSchLev-JCTC-24} These capabilities position OO-DFT as both a practical tool and a compelling alternative to (linear-response) time-dependent DFT (TDDFT) \cite{RunGro-PRL-84,AppGroBur-PRL-03,BurWerGro-JCP-05} for these types of excitations. \cite{TozRogHanRooSer-MP-99,TozHan-PCCP-00,DreWelHea-JCP-03,MaiZhaCavBur-JCP-04,LevKoQueMar-MP-06,Mai-JPCM-17}
This growing interest has fueled recent efforts to further refine and formalize the mathematical foundation of OO-DFT. \cite{YanAye-arXiv-24,Gou-PRA-25,GouDalKroPit-arXiv-24,Fro-JPCA-25}

In OO-DFT, self-consistent equations are solved for a non-Aufbau determinant constructed from KS orbitals. 
Here, ``non-Aufbau'' refers to not strictly occupying the lowest-energy KS orbitals.
This approach enables a state-specific representation, leading to KS orbitals, orbital energies, and potentials that vary across electronic states.
Since no density-functional approximations (DFAs) have been specifically developed for excited states, these calculations often rely on DFAs originally designed for the ground state.
Recent efforts by Gould, Pittalis and collaborators have addressed this gap by developing a unified DFA for ground and excited states (GX24) \cite{GouDalKroPit-arXiv-24} and a local-density approximation designed specifically for excited states. \cite{GouPit-PRX-24}
These advancements represent a significant step forward in developing state-specific functionals but further improvements are clearly necessary.

In a recent paper, we found that a distinct universal functional exists for each (singlet) state of the asymmetric Hubbard dimer at half-filling. \cite{GiaLoo-JPCL-23}
Expanding upon our recent findings, we investigate the extension of the ground-state KS-DFT formalism to the excited states of the Hubbard dimer.
Here, we show that while the KS-DFT framework is readily adaptable to describe the highest doubly-excited state, addressing the first excited state proves considerably more challenging.
Using a density-fixed adiabatic connection, we demonstrate that the density of the first excited state lacks non-interacting $v$-representability.
However, by introducing an analytic continuation of the adiabatic path, we show that the first excited state density becomes non-interacting complex-$v$-representable, as it can be reproduced via a complex-valued external potential.
From a practical perspective, we also perform state-specific KS calculations employing both exact and approximate correlation functionals. 
This reveals that approximate functionals can lead to spurious stationary solutions of the KS equations, highlighting key characteristics and limitations of this methodology, and suggesting the need for the development of state-specific functionals. 

Note that in the work of Ayers and Yang, \cite{YanAye-arXiv-24} it is found that the \textit{same} exchange-correlation functional is used by ground and excited states. 
However, their conclusion applies to functionals that are explicit in either one of the following variables: the KS state, the corresponding one-body reduced density matrix, or the excitation quantum number \textit{and} the KS potential.
By contrast, in this work, we consider state-specific functionals that depend directly on the density. 

%%%%%%%%%%%%%%%%%%%%%%%%%%%%%%
\section{Universal Density Functional}
\label{sec:UDF}
%%%%%%%%%%%%%%%%%%%%%%%%%%%%%%

%========================%
\subsection{Ground State}
\label{sec:UDF_GS}
%========================%
Here, we review the ground-state theory. \cite{Esc-Book-1996,Kut-JMS-06,Tou-SIP-23}
Below, we define the universal (i.e., independent of the external potential) ground-state functional of the density $F^{(0)}[\rho]$ using Levy’s constrained search \cite{Lev-PNAS-79} and Lieb’s convex formulation, \cite{Lie-IJQC-83} two complementary approaches that offer mathematical rigor and generality to DFT.

Within the Levy constrained search, \cite{Lev-PNAS-79} it can be defined as
\begin{equation}\label{eq:F0LL}
\begin{split}
	F^{(0)}[\rho] 
	& = \min_{\substack{\Psi \\ \rho_\Psi = \rho}} \mel{\Psi}{\hT + \hW}{\Psi}
	\\
	& = \mel*{\Psi^{(0)}[\rho]}{\hT + \hW}{\Psi^{(0)}[\rho]}
\end{split}
\end{equation}
with $\hT$ the kinetic energy operator and $\hW$ the electron repulsion operator. 
The minimization (which is independent of the external potential) is performed over all possible $N$-electron antisymmetric wave functions $\Psi$ that yield a given electron density $\rho$, i.e., $\rho_\Psi = \rho$.
The minimizing wave function, denoted as $\Psi^{(0)}[\rho]$, is known to exist but may not be unique. \cite{Lie-IJQC-83}
The exact ground-state energy is thus
\begin{equation} \label{eq:E0}
	E^{(0)}[v] = \min_{\rho} \qty{ F^{(0)}[\rho] +  \int v(\br) \rho(\br) \dd\br }
\end{equation}
where $v(\br)$ is the external local potential and $\int \rho(\br) \dd \br = N$.
The minimizer 
\begin{equation} \label{eq:rho0}
	\rho^{(0)}(\br) = \arg\min_{\rho} \qty{ F^{(0)}[\rho] +  \int v(\br) \rho(\br) \dd\br }
\end{equation}
is known to be the ground-state density if it exists.

Alternatively, by adopting the Lieb formulation, \cite{Lie-IJQC-83} the universal density functional is
\begin{equation}\label{eq:F0L}
	F^{(0)}[\rho] 
	= \max_{v}\qty{E^{(0)}[v] - \int v(\br) \rho(\br) \dd\br }
\end{equation}
with 
\begin{equation} \label{eq:E0min}
	E^{(0)}[v] = \min_\Psi \mel{\Psi}{\hH[v]}{\Psi}
\end{equation}
and
\begin{equation}\label{eq:H}
	\hH[v]= \hT + \hW + \hV[v]
\end{equation}
where $\hV[v] = \sum_i^N v(\br_i)$.
In Eq.~\eqref{eq:F0L}, the maximization is carried out over the external potential $v$ for a fixed density $\rho$, whereas the energy [see Eq.~\eqref{eq:E0min}] is evaluated at $v$ but may correspond to a different density.
We refer the interested reader to Ref.~\onlinecite{LewLieSei-SIP-23} for an exhaustive discussion about the distinct features of these universal functionals.

If they exist, the minimizer defined in Eq.~\eqref{eq:rho0} and the maximizer
\begin{equation}\label{eq:v0}
	v^{(0)}(\br) 
	= \arg\max_{v}\qty{E^{(0)}[v] - \int v(\br) \rho(\br) \dd\br }
\end{equation}
are both functionals of $\rho$ and are linked via the following relationships
\begin{subequations} 
\begin{align} 
	\label{eq:v0_vs_F0}
	v^{(0)}(\br) & = - \eval{\fdv{F^{(0)}[\rho(\br)]}{\rho(\br)}}_{\rho = \rho^{(0)}}
	\\
	\label{eq:rho0_vs_E0}
	\rho^{(0)}(\br) & = + \eval{\fdv{E^{(0)}[v(\br)]}{v(\br)}}_{v = v^{(0)}}
\end{align}
\end{subequations}
which shows that the energy and the universal functional are conjugate functions while $v$ and $\rho$ are dual quantities.
Note that here we assume differentiability although $F^{(0)}[\rho]$ is only ``almost differentiable'' which means that it may be approximated to any accuracy by a differentiable regularized functional. \cite{Lam-PRA-10,KvaEksTeaHel-JCP-14,LaePenTekRug-JCP-18,LaeTelPenRug-JCTC-19,PenLaeTelRug-PRL-19,ShiWas-JPCL-21,HelTea-EMS-22,PenCsiLae-ES-23,Gou-JCP-23,HerBakKae-arXiv-24}

%=========================%
\subsection{Excited States}
\label{sec:UDF_ES}
%=========================%
In his seminal paper, \cite{Gor-PRA-99} G\"orling proposes to replace the minimization in Levy's constrained search [see Eq.~\eqref{eq:F0LL}] by a general optimization procedure based on the stationarity principle, \cite{PerLev-PRB-85,Gor-PRA-96} where one seeks the stationary points of $\mel{\Psi}{\hT + \hW}{\Psi}$ with respect to $\Psi$ (see also Ref.~\onlinecite{YanAye-arXiv-24,Gou-PRA-25,GouDalKroPit-arXiv-24,Fro-JPCA-25}): 
\begin{equation} \label{eq:FmLL}
\begin{split}
	F^{(m)}[\rho] 
	& = \stat_{\substack{\Psi \\ \rho_\Psi = \rho}}\mel{\Psi}{\hT + \hW}{\Psi}
	\\
	& = \mel*{\Psi^{(m)}[\rho]}{\hT+ \hW}{\Psi^{(m)}[\rho]}
\end{split}
\end{equation}
where the subscript $m$ indexes the different stationary states $\Psi^{(m)}$, $m=0$ being the ground state.
% that we shall omit in the following.
The exact energy of the $m$th excited state can be expressed as
\begin{equation} \label{eq:Em}
	E^{(m)}[v] = \stat_{\rho} \qty{ F^{(m)}[\rho] +  \int v(\br) \rho(\br) \dd\br }
\end{equation}
An analogous generalization of the Lieb formulation [see Eq.~\eqref{eq:F0L}] yields
\begin{equation} \label{eq:FmL}
	F^{(m)}[\rho] 
	= \stat_{v} \qty{ E^{(m)}[v] - \int v(\br) \rho(\br) \dd\br }
\end{equation}
with 
\begin{equation}
	E^{(m)}[v] = \stat_{\Psi} \mel{\Psi}{\hH[v]}{\Psi}
\end{equation}
Optimizers (assuming they exist) of Eqs.~\eqref{eq:Em} and \eqref{eq:FmL} are given by \cite{Gou-PRA-25,Fro-JPCA-25}
\begin{subequations} 
\begin{align} 
	\label{eq:rhom}
	\rho^{(m)}(\br) 
	& = \arg\stat_{\rho} \qty{ F^{(m)}[\rho] +  \int v(\br) \rho(\br) \dd\br }
	\\
	\label{eq:vm}
	v^{(m)}(\br) 
	& = \arg\stat_{v}\qty{E^{(m)}[v] - \int v(\br) \rho(\br) \dd\br }
\end{align}
\end{subequations}

In our recent work, \cite{GiaLoo-JPCL-23} we applied the stationarity principle as in Eq.~\eqref{eq:FmLL} to obtain the universal functionals $F^{(m)}[\rho]$ for the excited states of a model system. 
We also applied an analogous stationarity principle on the Lieb formulation [see Eq.~\eqref{eq:FmL}] obtaining identical $F^{(m)}[\rho]$ as from the Levy constrained search.
Moreover, we showed that Eqs.~\eqref{eq:v0_vs_F0} and \eqref{eq:rho0_vs_E0} apply to the singlet excited states as well as to the ground state of the model, that is,
\begin{subequations} 
\begin{align} 
	\label{eq:vm_vs_Fm}
	v^{(m)}(\br) & = - \eval{\fdv{F^{(m)}[\rho]}{\rho(\br)}}_{\rho = \rho^{(m)}}
	\\
	\label{eq:rhom_vs_Em}
	\rho^{(m)}(\br) & = + \eval{\fdv{E^{(m)}[v]}{v(\br)}}_{v = v^{(m)}}
\end{align}
\end{subequations}
Although generalizations have been attempted in Refs.~\onlinecite{Gou-PRA-25,Fro-JPCA-25}, a general proof has not yet been established.

%%%%%%%%%%%%%%%%%%%%%%%%%%
\section{Kohn-Sham Scheme}
\label{sec:KS}
%%%%%%%%%%%%%%%%%%%%%%%%%%

%========================%
\subsection{Ground State}
\label{sec:KS_GS}
%========================%
As mentioned in Sec.~\ref{sec:intro}, the KS system corresponds to a system of (formally) non-interacting electrons. \cite{KohSha-PRA-65} 
For this special case, the universal functional is often referred to as $\Ts[\rho]$, the KS kinetic-energy functional. 
Within the constrained-search formalism, the non-interacting kinetic-energy functional is expressed as
\begin{equation}
\begin{split}
	\Ts^{(0)}[\rho] 
	& = \min_{\substack{\Phi \\ \rho_\Phi = \rho}} \mel{\Phi}{\hT}{\Phi}
	\\
	& = \mel*{\Phi^{(0)}[\rho]}{\hT}{\Phi^{(0)}[\rho]}
\end{split}
\end{equation}
where the minimization is performed, for non-degenerate ground states, over all possible $N$-electron single-determinant wave functions $\Phi$ that yield the fixed density $\rho$. 
Again, the minimizing wave function, denoted as $\Phi^{(0)}[\rho]$ and named KS wave function, is known to exist but may not be unique. \cite{Lie-IJQC-83}

The universal functional and the KS kinetic energy functional are linked by the Hartree-exchange-correlation (Hxc) functional, as follows:
\begin{equation}
	F^{(0)}[\rho] = \Ts^{(0)}[\rho] + \Ehxc^{(0)}[\rho] 
\end{equation}

Thanks to the introduction of the KS fictitious system, one can recast the minimization over densities in Eq.~\eqref{eq:E0} by a minimization over single-determinant wave functions %\Sara{Eq.~\eqref{eq:E0_CS} does not hold}
\begin{equation} \label{eq:E0_CS}
	E^{(0)}[v] = \min_{\Phi} \qty{ \mel{\Phi}{\hT+\hV[v]}{\Phi} + \Ehxc^{(0)}[\rho_\Phi]}
\end{equation}
where $\rho_\Phi$ is the density produced by $\Phi$. 
If the minimizing single-determinant wave function $\Phi^{(0)}$ exists, then it yields the exact ground-state density $\rho^{(0)}(\br)$.

The Hxc functional is often decomposed as
\begin{equation}\label{eq:Ehxc0}
	\Ehxc^{(0)}[\rho] = \Eh[\rho] + \Ex^{(0)}[\rho] + \Ec^{(0)}[\rho] 
\end{equation}
where
\begin{subequations}
\begin{align}
	\label{eq:Eh}
	\Eh[\rho] & = \frac{1}{2} \iint \frac{\rho(\br)\rho(\br')}{\abs{\br-\br'}} \dd\br\dd\br'
	\\
	\label{eq:Ehx}
	\Ex^{(0)}[\rho] & = \mel*{\Phi^{(0)}[\rho]}{\hW}{\Phi^{(0)}[\rho]} - \Eh[\rho]
	\\
	\label{eq:Ec}
	\begin{split}
		\Ec^{(0)}[\rho] & = \mel*{\Psi^{(0)}[\rho]}{\hT+\hW}{\Psi^{(0)}[\rho]} 
		\\
		& - \mel*{\Phi^{(0)}[\rho]}{\hT+\hW}{\Phi^{(0)}[\rho]}
	\end{split}
\end{align}
\end{subequations}
are the Hartree (H), exchange (x), and correlation (c) energy functionals, respectively.
The correlation part $\Ec^{(0)}[\rho] = \Tc^{(0)}[\rho] + \Wc^{(0)}[\rho]$ is the sum of a kinetic contribution 
\begin{equation}
	\Tc^{(0)}[\rho] = \mel*{\Psi^{(0)}[\rho]}{\hT}{\Psi^{(0)}[\rho]} - \mel*{\Phi^{(0)}[\rho]}{\hT}{\Phi^{(0)}[\rho]}
\end{equation} 
and a potential contribution
\begin{equation}
	\Wc^{(0)}[\rho] = \mel*{\Psi^{(0)}[\rho]}{\hW}{\Psi^{(0)}[\rho]} - \mel*{\Phi^{(0)}[\rho]}{\hW}{\Phi^{(0)}[\rho]}
\end{equation} 

According to KS-DFT, the ground-state density of the interacting Hamiltonian $\hH[v]$ can be obtained from 
\begin{equation}\label{eq:KSdens}
	\rho^{(0)}(\br) = \sum_{i} f_i^{(0)} \abs*{\psi_i^{(0)}(\br,\sigma)}^2
\end{equation}
where $f_i^{(0)}$ is the occupancy of the ground-state KS orbital $\psi_i^{(0)}(\br,\sigma)$ with spin $\sigma \in \qty{\uparrow,\downarrow}$.
In the ground-state case, we have
\begin{equation}
	f_i^{(0)} = 
	\begin{cases}
		1	\qif{i \le N}
		\\
		0	\qif{i > N}
	\end{cases}
\end{equation}
which corresponds to populate the $N$ lowest-energy KS orbitals (Aufbau filling).
The ground-state KS orbitals are obtained as the solutions to a set of single-particle equations, known as the KS equations: \cite{KohSha-PRA-65} 
\begin{equation}\label{eq:KS0eqs}
	\qty[ - \frac{\nabla_{\br}^2}{2} + v(\br) + \vhxc^{(0)}(\br)] \psi_i^{(0)}(\br) = \epsilon_i^{(0)} \psi_i^{(0)}(\br)
\end{equation}
where the $\epsilon_i^{(0)}$'s are KS orbital energies and the Hxc potential is
\begin{equation} \label{eq:vhxc0}
	\vhxc^{(0)}(\br) = \fdv{\Ehxc^{(0)}[\rho(\br)]}{\rho(\br)} 
\end{equation}
which can be decomposed into individual contributions, $\vhxc^{(0)}(\br) = \vh(\br) + \vx^{(0)}(\br) + \vc^{(0)}(\br)$, and the ground-state KS potential is, up to an additive constant, $\vs^{(0)}(\br) = v(\br) + \vhxc^{(0)}(\br)$ or
\begin{equation} \label{eq:ts0}
	\vs^{(0)}(\br) = - \fdv{\Ts^{(0)}[\rho(\br)]}{\rho(\br)}.
\end{equation}
Since $\vhxc^{(0)}$ is itself a functional of the density, the problem becomes non-linear and must be solved self-consistently.
Under certain assumptions (the desired density should be both interacting and non-interacting $v$-representable) and if the exact expression of $\Exc[\rho]$ were known (in an appropriate domain of $\rho$), Eq.~\eqref{eq:KSdens} would yield exactly the density of the interacting system.

It is possible to link the physical and KS systems by performing an adiabatic connection while keeping the ground-state density fixed. \cite{GunLun-PRB-76,LanPer-SSC-75,TeaCorHel-JCP-10} 
In practice, it is done by linearly scaling the electron-electron interaction with a coupling constant $\lambda \in \mathbb{R}$.
The Hamiltonian is thus defined as 
\begin{equation}\label{eq:H0_lambda}
	\hH_\lambda[v_\lambda^{(0)}] = \hT + \lambda \hW + \hV[v_\lambda^{(0)}]
\end{equation}
where $v_\lambda^{(0)}$ is the external potential that imposes that the ground-state density remains constant along the adiabatic path between the physical system at $\lambda = 1$ and the KS system at $\lambda = 0$.
In this setting, the universal ground-state functional along the adiabatic path is defined as 
\begin{equation}\label{eq:F0LL_lambda}
\begin{split}
	F_\lambda^{(0)}[\rho] 
	& = \min_{\substack{\Psi \\ \rho_\Psi = \rho}} \mel{\Psi}{\hT + \lambda \hW}{\Psi}
	\\
	& = \mel*{\Psi_\lambda^{(0)}[\rho]}{\hT + \lambda \hW}{\Psi_\lambda^{(0)}[\rho]}
\end{split}
\end{equation}
or equivalently
\begin{equation}\label{eq:F0L_lambda}
	F_\lambda^{(0)}[\rho] 
	= \max_{v}\qty{E_\lambda^{(0)}[v] - \int v(\br) \rho(\br) \dd\br }
\end{equation}
which takes the following limiting values at the extremities of the path
\begin{align}
	F_{\lambda=0}^{(0)}[\rho] & = \Ts^{(0)}[\rho] 
	& 
	F_{\lambda=1}^{(0)}[\rho] & = F^{(0)}[\rho]
\end{align}

%=========================%
\subsection{Excited States}
\label{sec:KS_ES}
%=========================%
Here, we generalized the ground-state KS scheme described in Sec.~\ref{sec:UDF_GS} to excited states. \cite{Gor-PRA-96,EvaShuTul-JPCA-13,YanAye-arXiv-24}
First, thanks to the stationarity principle developed in Sec.~\ref{sec:UDF_ES}, we generalize the non-interacting kinetic-energy functional to excited states, such that
\begin{equation}
\begin{split}
	\Ts^{(m)}[\rho] 
	& = \stat_{\substack{\Psi \\ \rho_\Psi = \rho}} \mel{\Phi}{\hT}{\Phi}
	\\
	& = \mel*{\Phi^{(m)}[\rho]}{\hT}{\Phi^{(m)}[\rho]}
\end{split}
\end{equation}
where the search of stationary points is performed over the set of $N$-electron single-determinant wave functions with potentially non-Aufbau fillings.
We note that the KS wave function is not necessarily restricted to a single Slater determinant; in the degenerate case, it can be expressed as a linear combination of Slater determinants. \cite{Lie-IJQC-83,Tou-SIP-23}
(The same comment applies to the ground state.)

There might exist more than one single-determinant wave function with a different index $m$ that yields the fixed density $\rho$.
Although we use the same index ($m$) for the KS and exact excited states, it is important to mention that the rankings in the KS and exact frameworks are not necessarily identical.
In other words, $m$ should be regarded more as a tracking number than a ranking number.

Within the KS scheme, the universal functional associated with the $m$th excited state is thus decomposed as
\begin{equation}\label{eq:F_KS}
	F^{(m)}[\rho] = \Ts^{(m)}[\rho] + \Ehxc^{(m)}[\rho] 
\end{equation}
and the corresponding energy reads 
\begin{equation} \label{eq:Em_CS}
	E^{(m)}[v] = \stat_{\Phi} \qty{ \mel{\Phi}{\hT+\hV[v]}{\Phi} + \Ehxc^{(m)}[\rho_\Phi]}
\end{equation}
The state-specific Hxc functional can be decomposed as
\begin{equation}\label{eq:Ehxc_mu}
	\Ehxc^{(m)}[\rho] = \Eh[\rho] + \Ex^{(m)}[\rho] + \Ec^{(m)}[\rho] 
\end{equation}
where the Hartree functional given by Eq.~\eqref{eq:Eh} is the only state-independent quantity, the exchange functional being
\begin{equation}
	\Ex^{(m)}[\rho] = \mel*{\Phi^{(m)}[\rho]}{\hW}{\Phi^{(m)}[\rho]} - \Eh[\rho]
\end{equation}
while the correlation functional is
\begin{multline}
	\Ec^{(m)}[\rho] 
	= \mel*{\Psi^{(m)}[\rho]}{\hT+\hW}{\Psi^{(m)}[\rho]} 
	\\
	- \mel*{\Phi^{(m)}[\rho]}{\hT+\hW}{\Phi^{(m)}[\rho]}
\end{multline}
Again, we split the correlation part as $\Ec^{(m)}[\rho] = \Tc^{(m)}[\rho] + \Wc^{(m)}[\rho]$ with 
\begin{equation}
	\Tc^{(m)}[\rho] = \mel*{\Psi^{(m)}[\rho]}{\hT}{\Psi^{(m)}[\rho]} - \mel*{\Phi^{(m)}[\rho]}{\hT}{\Phi^{(m)}[\rho]}
\end{equation}
and 
\begin{equation}
	\Wc^{(m)}[\rho] = \mel*{\Psi^{(m)}[\rho]}{\hW}{\Psi^{(m)}[\rho]} - \mel*{\Phi^{(m)}[\rho]}{\hW}{\Phi^{(m)}[\rho]}
\end{equation}

The excited-state density reads
\begin{equation}
	\rho^{(m)}(\br) = \sum_{i} f_i^{(m)} \abs*{\psi_i^{(m)}(\br,\sigma)}^2
\end{equation}
where non-Aufbau fillings are considered and the occupation numbers $f_i^{(m)}$ are excited-state-dependent.
The KS orbitals are obtained as the solutions to the state-specific KS equations:
\begin{equation}\label{eq:KSmueqs}
	\qty[ - \frac{\nabla_{\br}^2}{2} + \vs^{(m)}(\br)] \psi_i^{(m)}(\br) = \epsilon_i^{(m)} \psi_i^{(m)}(\br)
\end{equation}
where the excited-state KS potential is given by $\vs^{(m)}(\br) = v(\br) + \vhxc^{(m)}(\br)$ or 
\begin{equation} \label{eq:ts}
	\vs^{(m)}(\br) = - \fdv{\Ts^{(m)}[\rho(\br)]}{\rho(\br)} 
\end{equation}
and the state-specific Hxc potential is
\begin{align} \label{eq:vhxc}
	\vhxc^{(m)}(\br) & = \fdv{\Ehxc^{(m)}[\rho(\br)]}{\rho(\br)}
\end{align}
which can be decomposed into individual contributions: $\vhxc^{(m)}(\br) = \vh(\br) + \vx^{(m)}(\br) + \vc^{(m)}(\br)$.
Similarly to the ground-state formalism, we defined a density-fixed adiabatic connection for each excited state as
\begin{equation}\label{eq:FmLL_lambda}
\begin{split}
	F_\lambda^{(m)}[\rho] 
	& = \stat_{\substack{\Psi \\ \rho_\Psi = \rho}} \mel{\Psi}{\hT + \lambda \hW}{\Psi}
	\\
	& = \mel*{\Psi_\lambda^{(m)}[\rho]}{\hT + \lambda \hW}{\Psi_\lambda^{(m)}[\rho]}
\end{split}
\end{equation}
or equivalently
\begin{equation} \label{eq:FmL_lambda}
	F_\lambda^{(m)}[\rho] 
	= \stat_{v} \qty{ E_\lambda^{(m)}[v] - \int v(\br) \rho(\br) \dd\br }
\end{equation}
with
\begin{align}
	F_{\lambda=0}^{(m)}[\rho] & = \Ts^{(m)}[\rho] 
	& 
	F_{\lambda=1}^{(m)}[\rho] & = F^{(m)}[\rho]
\end{align}
In this case, the Hamiltonian along the adiabatic connection is
\begin{equation}\label{eq:Hm_lambda}
	\hH_\lambda[v_\lambda^{(m)}] = \hT + \lambda \hW + \hV[v_\lambda^{(m)}]
\end{equation}
where $v_\lambda^{(m)}$ imposes that the $m$th excited-state density remains constant along the adiabatic path between the $m$th physical excited-state and its corresponding KS state (if it exists).

%%%%%%%%%%%%%%%%%%%%%%%%%%%%%%%%%%
\section{Asymmetric Hubbard Dimer}
%%%%%%%%%%%%%%%%%%%%%%%%%%%%%%%%%%

%%%%%%%%%%%%%%%%%%%%%%%%%%%%%%%%%%
\subsection{Hamiltonian}
\begin{figure}
\includegraphics[width=0.6\linewidth]{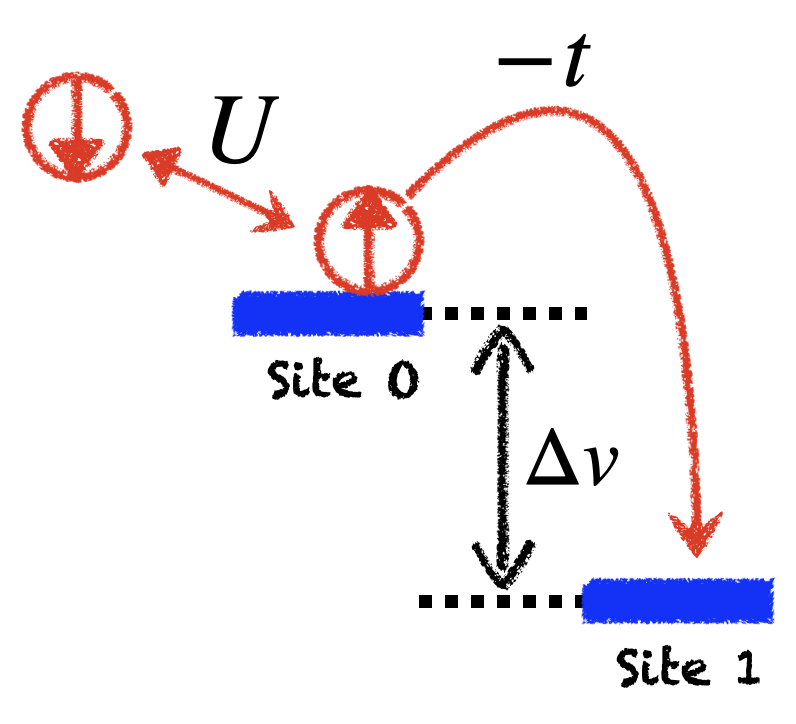}
\caption{Schematic representation of the asymmetric Hubbard dimer at half filling: $t$ is the hopping parameter, $U$ is the on-site interaction parameter, and $\Dv$ is the difference in potential between the two sites.}
	 \label{fig:Hubbard}
\end{figure}

%%%%%%%%%%%%%%%%%%%%%%%%%%%%%%%%%%
In the Hubbard dimer at half-filling (see Fig.~\ref{fig:Hubbard}), the Hamiltonian is given by \cite{Hub-PRSL-63,LieWu-PRL-68,SchGun-JPC-87,Mon-WS-92,CarFerBur-JPCM-15,CohMor-PRA-16,YinBroLopVarGorLor-PRB-16,SmiPriBur-PRB-16,SenTsuRobFro-MP-17,DeuMazFro-PRB-17,CarFerMaiBur-EPJB-18,GiaPri-JCP-22,QinSchAndCorGul-ARCMP-22,GiaPri-JCTC-23} 
\begin{equation}\label{eq:cH}
    \mathcal{\hH} = \mathcal{\hT} + \mathcal{\hat{W}} + \mathcal{\hV}
\end{equation}
with
\begin{subequations}
\begin{align}
	\mathcal{\hT} & = - t \sum_{\sigma=\up,\dw} \qty( \cre{0\sigma} \ani{1\sigma} + \text{h.c.} )
	\\
	\mathcal{\hW} & = U \sum_{i=0}^{1} \hat{n}_{i\up} \hat{n}_{i\dw}
	\\
	\mathcal{\hV} & = \Dv \frac{\hn_{1} - \hn_{0}}{2}
\end{align}
\end{subequations}
where $t > 0$ is the hopping parameter, $U \ge 0$ is the on-site interaction parameter, $\hn_{i\sigma} = \cre{i\sigma}\ani{i\sigma}$ is the spin-dependent site-occupation operator, $\hn_{i} = \hn_{i\uparrow} + \hn_{i\downarrow}$ is the spin-summed site-occupation operator, and $\Dv = v_{1}-v_{0}$ (with $v_0 + v_1 = 0$) is the potential difference (asymmetry) between the two sites. 
Note that because the operator $\mathcal{\hT}$ mimics in some sense the kinetic energy operator for quantum systems in real space, we will refer to its expectation value as ``kinetic energy'' despite some striking differences compared to the real-space case. 

%%% FIG 1 %%%

The total number of particles in the model is controlled by the sum of the expectation values of the occupation operator on each site, i.e., $  n_0+n_1= N $. 
Here, we consider only the case $N=2$ (half-filling). 
The site-occupation difference $\Dn =n_1-n_0 $ in the model plays the role of the electron probability density in real space and we have $-2 \le \Dn \le 2$. 
For notational convenience, we use the reduced quantity $\rho = \Dn/2$ throughout the paper.
Furthermore, we restrict ourselves to singlet states.
Although one technically deals with functions in the Hubbard dimer as $-1 \le \rho \le 1$, we shall stick to the term functional throughout the paper.

The Hamiltonian $\mathcal{\hH}$ lives in a three-dimensional Hilbert space which can be represented in the basis $\ket{ 0_\uparrow 0_\downarrow},\ket{1_\uparrow1_\downarrow} $, and $ \qty( \ket{0_\uparrow 1_\downarrow}-\ket{0_\downarrow 1_\uparrow} )/\sqrt{2}$ and has only three eigenvalues and eigenvectors, labeled  $E^{(m)}$ and $\Psi^{(m)}$, respectively.
Moreover, the general form of a singlet wave function is
\begin{equation}
	\ket{\Psi} = x \ket{0_\up0_\dw} + y \frac{\ket{0_\up1_\dw} - \ket{0_\dw1_\up}}{\sqrt{2}} + z \ket{1_\up1_\dw}
\end{equation}
with $-1 \le x,y,z \le 1$ and the normalization condition 
\begin{equation}
\label{eq:normalization}
	x^2 + y^2 + z^2 = 1
\end{equation}
The energy expression is simply $E = T + W + V$, with
\begin{subequations}
\begin{align}
	\label{eq:T}
	&T  = - 2\sqrt{2} t y \qty(x + z)
	\\
	\label{eq:Vee}
	&W  = U \qty(x^2 + z^2)
	\\
	\label{eq:V}
	&V  = \rho \, \Dv  
\end{align}
\end{subequations}
and
\begin{equation}
	\rho = z^2 - x^2
\end{equation}
Note that the normalized eigenstates of the model Hamiltonian are fully determined by only two parameters: $U/(2t)$ and $\Dv/(2t)$. 
For this reason, we set $t=1/2$ throughout the paper.

For finite and infinite lattice systems, Chayes, Chayes, and Ruskai proved that every ground-state density is $v$-representable. \cite{ChaChaRus-JSP-85}
However, the Hohenberg-Kohn theorem does not always hold (while a potential exists, it might not be unique). 
A counterexample has been recently discovered by Penz and van Leeuwen \cite{PenLee-JCP-23} and attributed to degeneracies. \cite{UllKoh-PRL-02,PenLee-Quantum-23,PenLee-JCP-24}
For Hubbard chains though, the uniqueness and thus the applicability of the Hohenberg-Kohn theorem are established for the ground state \cite{PenLee-JCP-23} but not much is known for excited states.

\begin{figure}
\includegraphics[width=\linewidth]{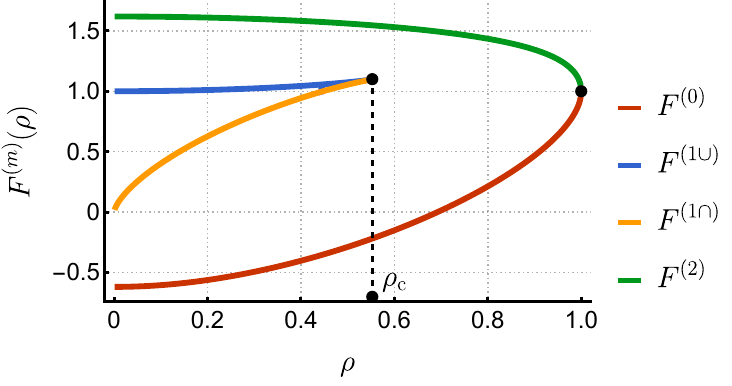}
\caption{Exact functional $F^{(m)}(\rho)$ as a function of $\rho$ for each singlet state of the asymmetric Hubbard dimer with $U=1$.}
	 \label{fig:F_vs_rho}
\end{figure}

Using Levy's constrained-search formulation \cite{Lev-PNAS-79} or, equivalently, Lieb's convex formulation, \cite{Lie-IJQC-83} we demonstrated in Ref.~\onlinecite{GiaLoo-JPCL-23} that the exact functionals corresponding to each singlet state of the asymmetric Hubbard dimer can be determined.
As illustrated in Fig.~\ref{fig:F_vs_rho}, the ground-state functional, $F^{(0)}(\rho)$, is convex with respect to $\rho$, while the doubly excited-state functional, $F^{(2)}(\rho)$, is concave.
Remarkably, the functional associated with the singly excited state is a partial (i.e., defined for a subdomain of $\rho$), multivalued functional that consists of one convex branch, $F^{(1\cup)}(\rho)$, and one concave branch, $F^{(1\cap)}(\rho)$, each associated with a separate domain of $\Dv$ values. 
This behavior arises from the non-invertibility of the density-potential mapping for the first excited state, where a single density $\rho$ can correspond to two different $\Dv$ values.

%%% FIG 2 %%%

%%%%%%%%%%%%%%%%%%%%%%%%%%%%%%%%%%
\subsection{Existence of an Adiabatic Connection}
%%%%%%%%%%%%%%%%%%%%%%%%%%%%%%%%%%
This section explores the hypothetical existence of a (non-interacting) KS reference state for each interacting state of the asymmetric Hubbard dimer at half-filling.
Specifically, we investigate the existence of a connection, at a fixed density, between the physical system at $\lambda = 1$ and the KS system at $\lambda = 0$.

To study this adiabatic connection, we rely on G\"orling's stationary principle as applied to Lieb's formulation. \cite{Lie-IJQC-83}
Within this formalism [see Eq.~\eqref{eq:FmL_lambda}], the exact functionals $F_\lambda^{(m)}(\rho)$ are obtained by seeking the set of stationary points with respect to $\Dv$ for each value of $m$ and strength $\lambda$, \ie,
\begin{equation} \label{eq:Fm_Hu}
	F_\lambda^{(m)}(\rho) =\stat_{\Dv}\qty[f_\lambda^{(m)}(\rho,\Dv)]
\end{equation}
with
\begin{equation}
\begin{split}
	f_\lambda^{(m)}(\rho,\Dv) 
	& = E_\lambda^{(m)} - \Dv \rho
	\\
	& = \frac{\mel*{\Psi_\lambda^{(m)}}{\mathcal{\hH}_\lambda }{\Psi_\lambda^{(m)}}}{\braket*{\Psi_\lambda^{(m)}}{\Psi_\lambda^{(m)}}} - \Dv \rho
\end{split}
\end{equation}
where the $E_\lambda^{(m)}$'s are given by the eigenvalues of the following Hamiltonian matrix
\begin{equation} \label{eq:cH_lambda}
	\mathcal{H}_\lambda 
	= \mqty(
		- \Dv & -\sqrt{2} t  & 0 \\
		-\sqrt{2} t & 0 & -\sqrt{2}t \\
		0 & -\sqrt{2}t & + \Dv \\
	) 
	+ \lambda \mqty(
		U  & 0  & 0 \\
		0 & 0 & 0 \\
		0 & 0 & U \\
	) 
\end{equation}
and $\Psi_\lambda^{(m)}$ is the exact wave function associated with the $m$th state at strength $\lambda$.
As indicated by Eq.~\eqref{eq:vm}, the external potential for a given $\rho$ is expressed as follows:
\begin{equation} 
	\Dv_\lambda^{(m)}(\rho) =\arg\stat_{\Dv}\qty[f_\lambda^{(m)}(\rho,\Dv)]
\end{equation}

%%% FIG 3 %%%
\begin{figure}
	\includegraphics[width=\linewidth]{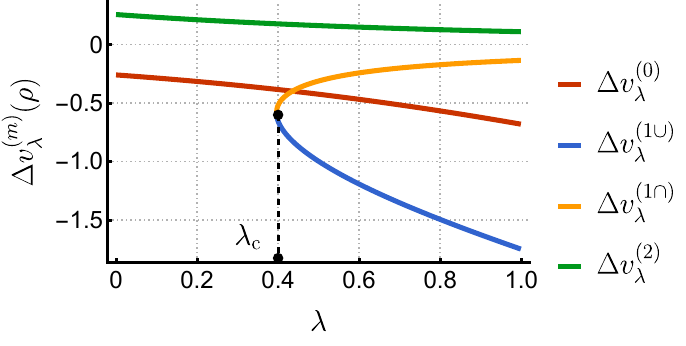}
	\includegraphics[width=\linewidth]{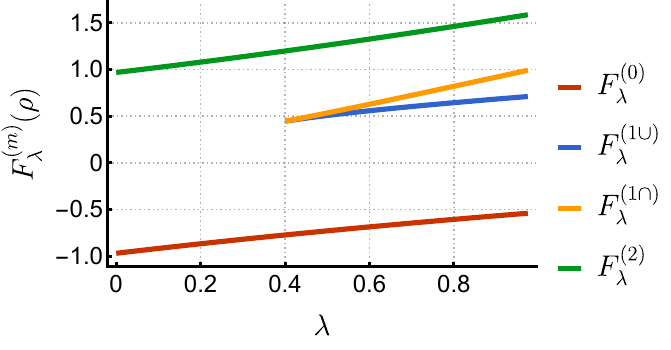}
	\includegraphics[width=\linewidth]{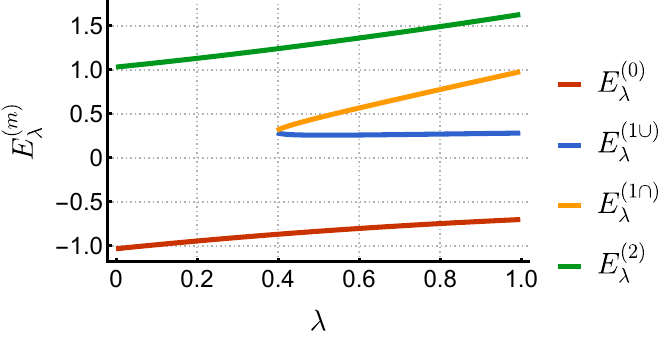}
	\caption{$\Dv_\lambda^{(m)}(\rho)$ (top), $F_\lambda^{(m)}(\rho)$ (center), and $E_\lambda^{(m)}(\rho)$ (bottom) as functions of $\lambda$ for $U = 1$ and $\rho=1/4$: ground state ($m=0$), singly-excited state ($m=1$), and doubly-excited state ($m=2$).
	$1\cup$ and $1\cap$ correspond to the convex and concave branches of the first excited state, which merge at $\lambda = \lambda_\text{c}$.}
	 \label{fig:Dv_vs_lambda}
\end{figure}

As shown in Fig.~\ref{fig:Dv_vs_lambda}, where we have depicted $\Dv_\lambda^{(m)}(\rho)$ (top), $F_\lambda^{(m)}(\rho)$ (center), and $E_\lambda^{(m)}(\rho)$ (bottom) as functions of $\lambda$ for $U = 1$ and $\rho=1/4$, the ground and doubly-excited states ($m = 0$ and $2$) are smoothly connected to their respective KS state. 
This connection holds for any $\rho$ and any $U$.
In contrast, the curves corresponding to the two branches of the first excited states ($1\cup$ and $1\cap$) merge at the critical value $\lambda = \lambda_\text{c}$ and disappear beyond this point. 
Therefore, at first glance, it would appear that there is no well-defined KS state for the first excited state as one cannot find an external potential that recreates this value of $\rho$ for $\lambda < \lambda_\text{c}$.
We can thus conclude that the first excited state of the Hubbard dimer lacks non-interacting $v$-representability.
%We note in passing that it can be readily shown that there is no issue of $v$-representability in the strictly-correlated limit, corresponding to $\lambda \to \infty$. 
%In the strictly-correlated limit, all densities are $v$-representable. \cite{FriGerGor-SIP-23}

Here, we propose to analytically continue this adiabatic connection into the complex plane by allowing for complex-valued $\Dv$. 
One key consequence of this extension is that the Hamiltonian $\mathcal{\hH}$ defined in Eq.~\eqref{eq:cH} becomes a complex
symmetric operator.
To accommodate this non-Hermitian nature, we redefine the inner product, replacing the standard scalar product with the so-called c-product \cite{MoiseyevBook,Rei-ARPC-82,Moi-PR-98} defined as $\cbraket{f}{g} = \braket{f^*}{g}$.
This symmetric bilinear form, while useful, does not constitute a valid metric norm, as the c-norm $\cbraket{f}{f}$ can become zero for non-zero functions $f$, a phenomenon known as self-orthogonality. 
Consequently, the eigenfunctions of $\mathcal{\hH}$ may not form a complete set.
Nonetheless, thanks to this generalized inner product, the energy can be computed for complex-valued $\Dv$ via the complex-stationary principle \cite{MoiCerWei-MP-78,BraFro-PRA-77,Moi-MP-82}
\begin{equation} \label{eq:stat_principle}
	E_\lambda^{(m)} = \frac{\cmel{\Psi_\lambda^{(m)}}{\mathcal{\hH_\lambda}}{\Psi_\lambda^{(m)}}}{\cbraket{\Psi_\lambda^{(m)}}{\Psi_\lambda^{(m)}}}
\end{equation}
Unlike the variational principle for Hermitian operators, Eq.~\eqref{eq:stat_principle} provides a stationarity condition for the complex energy without imposing bounds on its real or imaginary components. 
This extension of the domain of definition of the energy and the functional can be seen as a generalization of the holomorphic formalism \cite{HisTho-JCTC-14}
that allow stationary states to be analytically continued across molecular structures \cite{BurTho-JCTC-16,BurGroTho-JCTC-18} and exchange-correlation functionals. \cite{ZarBurTho-JCTC-20}
Here the analytic continuation is performed across the interaction strength at fixed density. \cite{BurThoLoo-JCP-19}
It also parallels the formalism introduced by Ernzerhof for open systems \cite{Ern-JCP-2006,ZhoErn-JCP-12} and the principles of density-functional resonance theory. \cite{WasMoi-PRL-07,WhiWas-JPCL-10,WhiWas-PRL-11,WhiWas-JCP-12}

%%% FIG 4 %%%
\begin{figure}
	\includegraphics[width=\linewidth]{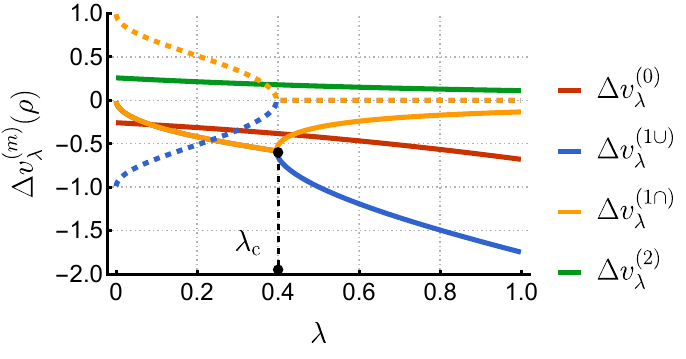}
	\includegraphics[width=\linewidth]{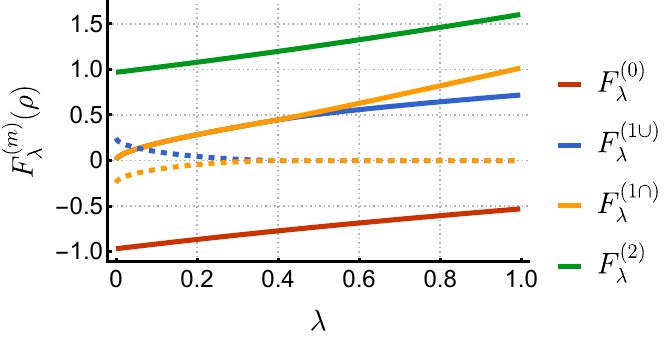}
	\includegraphics[width=\linewidth]{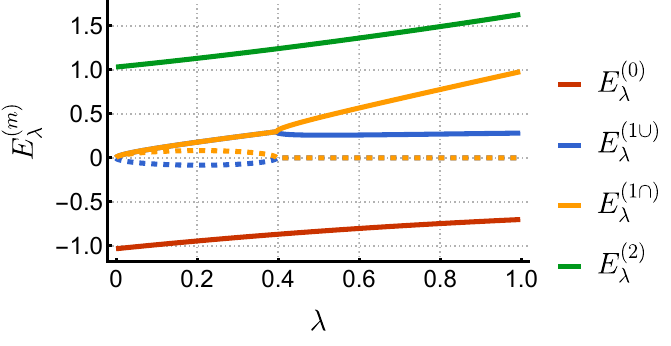}
	\caption{Analytic continuation of $\Dv_\lambda^{(m)}(\rho)$ (top), $F_\lambda^{(m)}(\rho)$ (center) and $E_\lambda^{(m)}$ (bottom) as functions of $\lambda$ for $U = 1$ and $\rho=1/4$: ground state ($m=0$), singly-excited state ($m=1$), and doubly-excited state ($m=2$). 
	$1\cup$ and $1\cap$ correspond to the convex and concave branches of the first excited state which become complex-valued for $\lambda < \lambda_\text{c}$. 
	The solid lines correspond to the real parts while the dashed lines represent the imaginary components.}
	 \label{fig:ReIm_vs_lambda}
\end{figure}

The top panel of Fig.~\ref{fig:ReIm_vs_lambda} illustrates the behavior of $\Dv_\lambda^{(m)}(\rho)$ as it evolves beyond the critical $\lambda$ value. 
After the merging point, the two stationary points transition into a pair of complex conjugate solutions and smoothly evolve from $\lambda = \lambda_\text{c}$ to $\lambda = 0$.
In the non-interacting limit ($\lambda = 0$), $\Dv_\lambda^{(m)}(\rho)$ is purely imaginary for the first excited state, whereas it has both real and imaginary components for $0 < \lambda < \lambda_\text{c}$.

The central and bottom panels of Fig.~\ref{fig:ReIm_vs_lambda} report the evolution of $F_\lambda^{(m)}(\rho)$ and $E_\lambda^{(m)}$, respectively, as functions of $\lambda$.
As expected, both quantities become complex-valued past the critical interaction strength. 
Interestingly, the energy is zero at $\lambda = 0$ for all $\rho$ values while the exact functional associated with the first excited state is purely imaginary.

Let us write down the various quantities at $\lambda = 0$ in a more explicit form.
By solving the following sixth-order polynomial equation
\begin{multline} \label{eq:pol}
	\rho^2 + \qty(3\rho^2 - 1) \qty(\frac{\Dv}{2t})^2 
	+ \qty(3\rho^2 - 2) \qty(\frac{\Dv}{2t})^4  
	\\
	+ \qty(\rho^2 - 1) \qty(\frac{\Dv}{2t})^6 = 0
\end{multline}
which is obtained as the limit of
\begin{equation}
	\pdv{f_\lambda^{(m)}(\rho,\Dv)}{\Dv} = 0
\end{equation}
as $\lambda$ goes to zero,
it is easy to show that the optimizers in the non-interacting limit are
\begin{subequations}
\begin{align}
	\Dv_0^{(0)}(\rho) & = \vs^{(0)}(\rho) = - \frac{2t \rho}{\sqrt{1 - \rho^2}} 
	\label{eq:vs0}
	\\
	\Dv_0^{(1\pm)}(\rho) & = \vs^{(1\pm)}(\rho)  =\pm 2t \ii   
	\label{eq:vs1}
	\\
	\Dv_0^{(2)}(\rho) & =\vs^{(2)}(\rho)  = + \frac{2t \rho}{\sqrt{1 - \rho^2}} 
	\label{eq:vs2}
\end{align}
\end{subequations}
where the two complex branches of the first excited states are denoted $1\pm$.
(The sixth-order polynomial defined in Eq.~\eqref{eq:pol} has two additional solutions that are purely imaginary for all values of $\rho$. At this point, their physical and mathematical significance remains unclear.)
From these quantities, we get the following non-interacting kinetic energies
\begin{subequations}
\begin{align}
	\label{eq:Ts0}
	F_0^{(0)}(\rho) & = \Ts^{(0)}(\rho) = - 2t \sqrt{1 - \rho^2}
	\\
	\label{eq:Ts1}
	F_0^{(1\pm)}(\rho) & = \Ts^{(1\pm)}(\rho)= \pm 2t \ii \rho
	\\
	\label{eq:Ts2}
	F_0^{(2)}(\rho) & = \Ts^{(2)}(\rho) = + 2t \sqrt{1 - \rho^2}
\end{align}
\end{subequations}
and the corresponding total energies read
\begin{subequations}
\begin{align}
	E_0^{(0)} & = - \frac{2t}{\sqrt{1 - \rho^2}}
	\\
	E_0^{(1)} & \equiv  E_0^{(1\pm)} = 0
	\\
	E_0^{(2)} & = + \frac{2t}{\sqrt{1 - \rho^2}}
\end{align}
\end{subequations}
We recover the well-known results for the ground state (see, for example, Ref.~\onlinecite{CarFerBur-JPCM-15}) and, as expected, the opposite values for the doubly-excited state.
Note that because the non-interacting kinetic energy is linear in $\rho$, the KS potential of the first excited state is independent of the density [see Eq.~\eqref{eq:ts}].%\Sara{???}
In the following, we arbitrarily assign the positive (negative) complex part to the concave (convex) branch.
Moreover, in the remaining, we omit the dependence in $\lambda$.

%%% FIG 5 %%%
\begin{figure}
\includegraphics[width=0.8\linewidth]{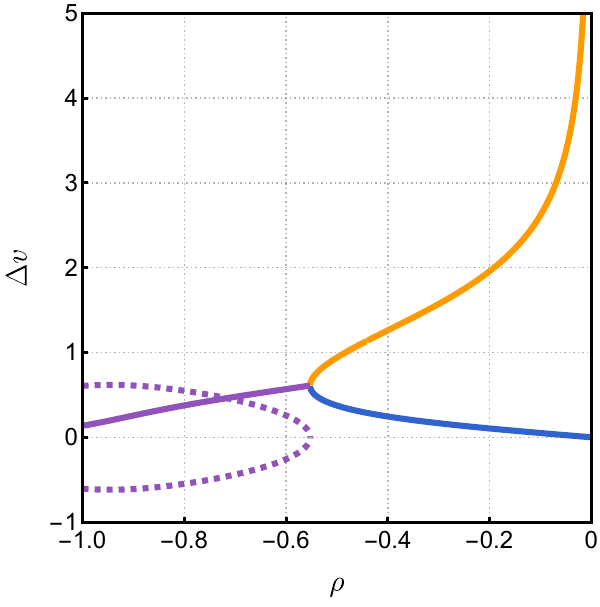}
\caption{Analytic continuation of $\Dv$ as a function of $\rho$ for the first excited state ($m=1$) for $U = 1$. The real and imaginary parts of $\Dv$ are represented as solid and dashed lines, respectively. The analytic continuation part is depicted in purple.}
	 \label{fig:Dv_vs_rho}
\end{figure}

To conclude this subsection, let us discuss the analytic continuation of the dual quantities $\Dv$ and $\rho$ for the fully-interacting system.
Figure \ref{fig:Dv_vs_rho} illustrates the analytic continuation of $\Dv$ as a function of $\rho$ (or vise versa) for the first excited state ($m=1$) for $U = 1$.
The analytic continuation is shown in purple, with the real part represented by a solid line and the imaginary parts depicted as dashed lines.
For a given value of $U$, densities higher than $\rho_\text{c}$ (in absolute value) cannot be reached with real-valued $\Dv$. 
Instead, values of $\rho$ exceeding $\rho_\text{c}$ are accessible only for complex-valued $\Dv$. 
For $\abs{\rho} < \abs{\rho_\text{c}}$, two distinct values of $\Dv$ correspond to each $\rho$: one on the convex branch for smaller $\Dv$ and another on the concave branch for larger $\Dv$.
%the density of the first excited state is interacting $v$-representable. 
%For the same range of $\rho$ values, as demonstrated above, a complex potential can generate this density.
% in the non-interacting case. 
%Thus, one could say that the first excited state densities are non-interacting complex-$v$-representable.
%Again, in the non-interacting case, 
Likewise, for $\abs{\rho} > \abs{\rho_\text{c}}$, it does not exist a unique $\Dv$ that generates this density but a pair of complex-conjugate values of the potential.
The density-potential map is thus non-invertible both on the real axis and in the complex plane.

%%%%%%%%%%%%%%%%%%%%%%%%%%%%%%%%%%
\section{Kohn-Sham in practice}
%%%%%%%%%%%%%%%%%%%%%%%%%%%%%%%%%%

\begin{figure}
\includegraphics[width=\linewidth]{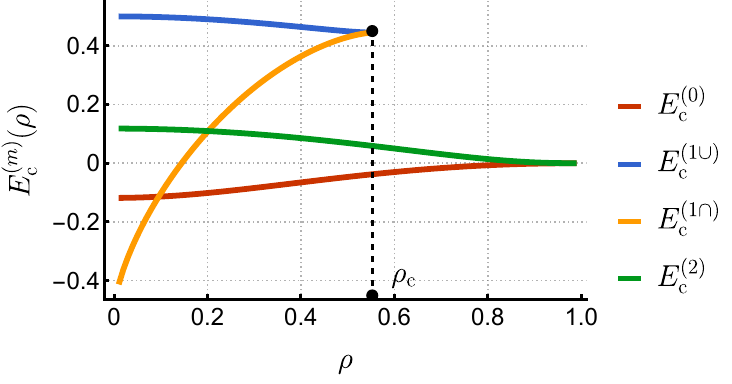}
\includegraphics[width=\linewidth]{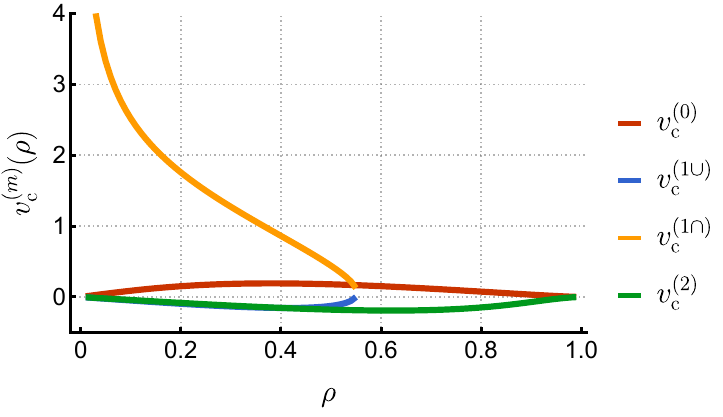}
\caption{$\Ec^{(m)}$ (top) and $\vc^{(m)}$ (bottom) as functions of $\rho$ for each singlet state of the asymmetric Hubbard dimer with $U=1$.}
	 \label{fig:Ec_vs_rho}
\end{figure}

We now perform state-specific KS calculations in practice using various Hxc functionals. 
In other words, we study the performance of the ground-state functional for excited-state calculations, and vice versa.
Ground-state and doubly-excited-state KS calculations are discussed in Secs.~\ref{sec:KS0} and \ref{sec:KS2}, respectively, while the singly-excited-state KS scheme is addressed in Sec.~\ref{sec:KS1}. 
Since the Hx contribution is identical for all states, the only distinction arises from the correlation component.
To illustrate this, Fig.~\ref{fig:Ec_vs_rho} depicts the correlation functional (top) and correlation potential (bottom) for each singlet state of the asymmetric Hubbard dimer.
As discussed below, these results reveal notable differences as well as intriguing similarities.

The correlation energy functionals are calculated as 
\begin{equation}
E_\text{c}^{(m)}(\rho) = F^{(m)}(\rho) - T_\text{s}^{(m)}(\rho) - E_\text{Hx} (\rho)
\end{equation}
where $F^{(m)}$ are the functionals shown in Fig.~\ref{fig:F_vs_rho} for each $m$, $T_\text{s}^{(m)}$ are given by Eqs.~\eqref{eq:Ts0}, \eqref{eq:Ts1}, and \eqref{eq:Ts2}, while $E_\text{Hx} = U/2(1+\rho^2)$ for any $m$.
The corresponding potentials are simply given by
$v_\text{c}^{(m)} (\rho) = \dv{E_\text{c}^{(m)}(\rho)}{\rho}$.
Finally, we note that for $m=1(\pm)$, we plot only the real part (compare Sec.~\ref{sec:KS1}).

%%% FIG 6 %%%

%%%%%%%%%%%%%%%%%%%%%%%%%%%%%%%%%%
\subsection{Ground-State Kohn-Sham Calculations}
\label{sec:KS0}
%%%%%%%%%%%%%%%%%%%%%%%%%%%%%%%%%%

In ground-state KS calculations, one seeks the values of $\rho$ such that the following equation is fulfilled:
\begin{equation} \label{eq:KS0}
	\vs^{(0)}(\rho) - v_\text{Hxc}^{(m)}(\rho) - \Dv = 0
\end{equation}
where $\vs^{(0)}$ is the ground-state KS potential [see Eq.~\eqref{eq:vs0}], $v_\text{Hxc}^{(m)} = U \rho + v_\text{c}^{(m)}$ is the Hxc potential of the $m$th state ($m = 0$, $1\cup$, $1\cap$, or $2$), and $\Dv$ is an input variable.
The corresponding KS energy is given by 
\begin{equation} \label{eq:EKS0}
	\EKS^{(0,m)} = \Ts^{(0)}(\rho^{(0,m)}) + E_\text{Hxc}^{(m)}(\rho^{(0,m)}) + \Dv \rho^{(0,m)}
\end{equation}
where $\Ts^{(0)}$ is given by Eq.~\eqref{eq:Ts0}, $E_\text{Hxc}^{(m)}$ is the Hxc functional of the $m$th state and $\rho^{(0,m)}$ is one of the roots of Eq.~\eqref{eq:KS0}.

%%% FIG 7 %%%
\begin{figure}
\includegraphics[width=\linewidth]{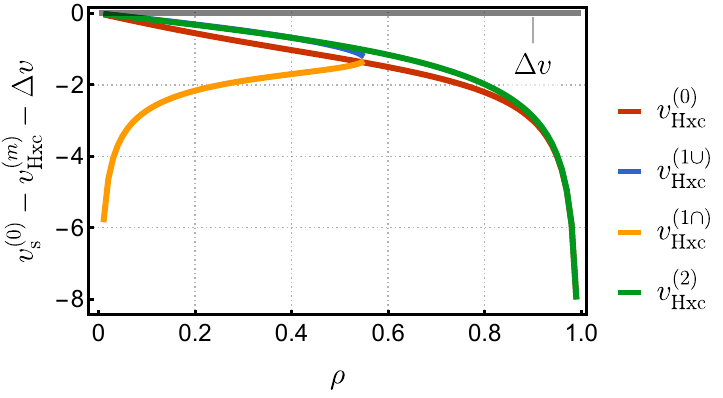}
\caption{$\vs^{(0)} - v_\text{Hxc}^{(m)} - \Dv$ as a function of $\rho$ for $U=1$. 
The horizontal gray line indicates the value of $\Dv$, which is set to zero here. 
Lowering $\Dv$ would move down this line.
For a given $\Dv$, the stationary densities are given by the intersection of the colored curves and the gray line.}
	 \label{fig:KSpot_Ts0}
\end{figure}

In Fig.~\ref{fig:KSpot_Ts0}, we plot $\vs^{(0)} - v_\text{Hxc}^{(m)} - \Dv$ as a function of $\rho$ for $U=1$ and $m = 0$ (ground state), $m = 1\cup$ (convex branch of the first excited state), $m = 1\cap$ (concave branch of the first excited state), and $m = 2$ (second excited state).
The horizontal gray line indicates the value of $\Dv$, which is set to zero here. 
Solutions for other values of $\Dv$ can be obtained by shifting down this gray line.
The stationary densities are given by the intersections of the colored curves and the gray line. 
The resulting stationary $\rho$ values may or may not be exact depending on the correlation potential (i.e., the value of $m$).

For $m = 0$ (red curve), one always gets a unique root that corresponds to the exact value of $\rho$ as $v_\text{Hxc}^{(0)}$ is the exact ground-state Hxc potential.
Using $v_\text{Hxc}^{(2)}$ instead (green curve), one gets a $\rho$ value quite close as generally $v_\text{Hxc}^{(2)}$ is a fairly good approximation of $v_\text{Hxc}^{(0)}$, especially for large $\Dv$. 
The convex branch of the first excited state (blue curve) is also a good approximation of $v_\text{Hxc}^{(0)}$ for $0 < \Dv < 1$.
Notably, there is a strong similarity between the curves obtained with $v_\text{Hxc}^{(1\cup)}$ and $v_\text{Hxc}^{(2)}$.
In all cases, a single solution is obtained across the range of $\Dv$, and these solutions correspond to minima.
This can be checked by computing the second derivative of the energy with respect to $\rho$, which is independent of $\Dv$.
Conversely, the concave branch (yellow curve) provides poor estimates of $\rho$ and these correspond to maxima rather than minima. 
This branch should be avoided for ground-state calculations.

%%%%%%%%%%%%%%%%%%%%%%%%%%%%%%%%%%
\subsection{Doubly-Excited-State Kohn-Sham Calculations}
\label{sec:KS2}
%%%%%%%%%%%%%%%%%%%%%%%%%%%%%%%%%%

In a doubly-excited state KS calculation, one looks for the zeros of the following equation: 
\begin{equation} \label{eq:ts2}
   \vs^{(2)}(\rho) - v_\text{Hxc}^{(m)}(\rho) - \Dv = 0 
\end{equation}
where $\vs^{(2)}$ is the KS potential associated with the second excited state [see Eq.~\eqref{eq:vs2}] and the expression of the KS energy reads
\begin{equation} \label{eq:EKS2}
	\EKS^{(2,m)} = \Ts^{(2)}(\rho^{(2,m)}) + E_\text{Hxc}^{(m)}(\rho^{(2,m)}) + \Dv \rho^{(2,m)}
\end{equation}
where $\Ts^{(2)}$ is given by Eq.~\eqref{eq:Ts2} and $\rho^{(2,m)}$ is a root of Eq.~\eqref{eq:ts2}.

%%% FIG 8 %%%
\begin{figure*}
\includegraphics[width=0.7\linewidth]{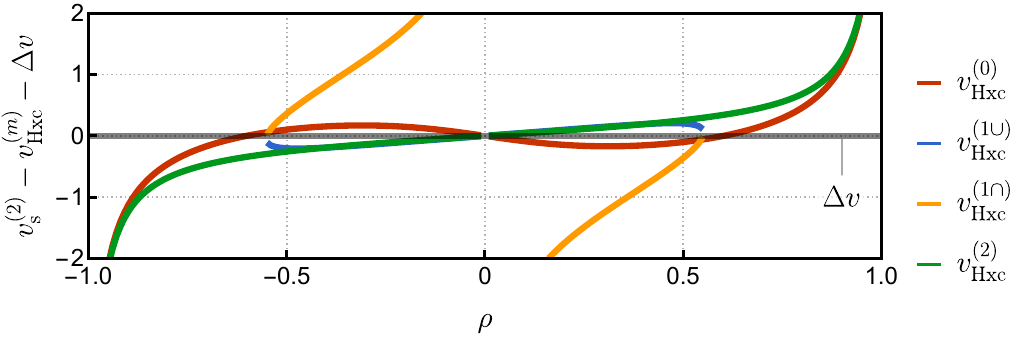}
\caption{$\vs^{(2)} - v_\text{Hxc}^{(m)} - \Dv$ as a function of $\rho$ for $U=1$. 
The horizontal gray line indicates the value of $\Dv$, which is set to zero here. 
Increasing (decreasing) $\Dv$ would move up (down) this line.
For a given $\Dv$, the stationary densities are given by the intersection of the colored curves and the gray line.}
	 \label{fig:KSpot_Ts2}
\end{figure*}

Analogously to Fig.~\ref{fig:KSpot_Ts0}, Fig.~\ref{fig:KSpot_Ts2} presents $\vs^{(2)} - v_\text{Hxc}^{(m)} - \Dv$ for both positive and negative values of $\rho$ as the non-monotonic nature of the potential makes things more interesting.
As before, the value of $\Dv$ is set to zero; increasing (decreasing) $\Dv$ shifts the line upward (downward).
Note that the stationary densities correspond to the intersections of the colored curves and the gray line.

As anticipated, $v_\text{Hxc}^{(2)}$ (green curve) yields the unique exact solution for any $\Dv$.
Note that these solutions are energy maxima.
Overall, $v_\text{Hxc}^{(1\cup)}$ (blue curve) is an excellent approximation of $v_\text{Hxc}^{(2)}$ for small $\Dv$ but its accuracy deteriorates near the junction with $v_\text{Hxc}^{(1\cap)}$, which, as in ground-state KS calculations (see Sec.~\ref{sec:KS0}), leads to extremely poor densities and energies.
The convex branch (yellow curve) should also be avoided for doubly-excited-state calculations.
$v_\text{Hxc}^{(0)}$ is a reasonable approximation of $v_\text{Hxc}^{(2)}$, particularly for large $\Dv$.
However, for small $\Dv$ ($\abs{\Dv} \lesssim 0.2$), using the ground-state functional for a doubly-excited state calculation can produce spurious solutions due to the presence of multiple roots.
Caution is therefore warranted, as convergence to a spurious solution (either a maximum or a minimum) can occur depending on the initial density. This issue becomes more pronounced at larger $U$, where correlation effects are stronger.

%%%%%%%%%%%%%%%%%%%%%%%%%%%%%%%%%%
\subsection{Singly-Excited-State Kohn-Sham Calculations}
\label{sec:KS1}
%%%%%%%%%%%%%%%%%%%%%%%%%%%%%%%%%%

\begin{figure}
\includegraphics[width=\linewidth]{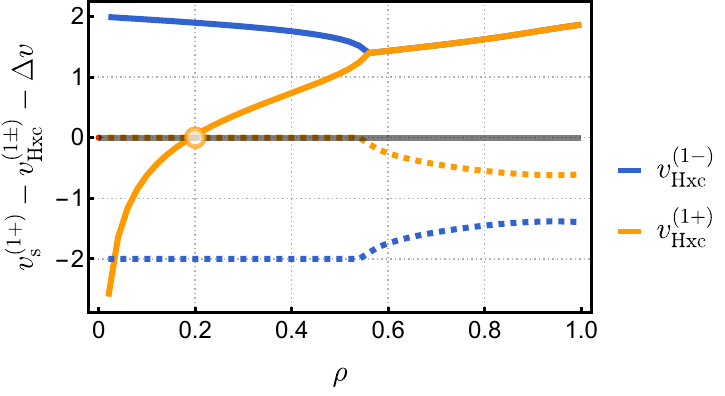}
\includegraphics[width=\linewidth]{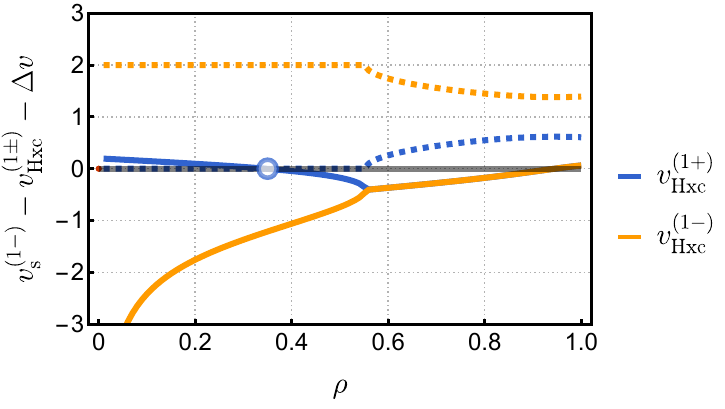}
\caption{$\vs^{(1+)} - v_\text{Hxc}^{(1\pm)} - \Dv$ (top) and $\vs^{(1-)} - v_\text{Hxc}^{(1\pm)} - \Dv$ (bottom) as functions of $\rho$ for $U=1$, $\Dv = -2$ (top), and $\Dv = -1/5$ (bottom). 
The zeros, as indicated by markers, correspond to the stationary self-consistent KS solutions.}
	 \label{fig:KSpot_Ts1pm}
\end{figure}

We now turn our attention to singly-excited-state KS calculations.
In this case, we seek the roots of the following complex-valued KS equation
\begin{equation} \label{eq:KS1pm}
	\vs^{(1\pm)} (\rho) - v_\text{Hxc}^{(m)}(\rho) - \Dv = 0
\end{equation}
where $m = 0$, $1-$, $1+$, or $2$, $\vs^{(1\pm)}$ is the KS potential associated with the first-excited state [see Eq.~\eqref{eq:vs1}] and the complex-valued KS energy is
\begin{equation} 
	\label{eq:EKS1pm}
	\EKS^{(1\pm,m)} = \Ts^{(1\pm)}(\rho^{(1\pm,m)}) + E_\text{Hxc}^{(m)}(\rho^{(1\pm,m)}) + \Dv \rho^{(1\pm,m)}
\end{equation}
Here, $\rho^{(1\pm,m)}$ is a real-valued root of Eq.~\eqref{eq:KS1pm}.
Complex-valued roots may exist but they are not considered here due to their unphysical nature.
Note that both the real and imaginary parts of Eq.~\eqref{eq:KS1pm} vanish at a stationary point, i.e., for $\rho = \rho^{(1\pm,m)}$.
This is illustrated in Fig.~\ref{fig:KSpot_Ts1pm} where we report $\vs^{(1+)} - v_\text{Hxc}^{(1\pm)} - \Dv$ (top) and $\vs^{(1-)} - v_\text{Hxc}^{(1\pm)} - \Dv$ (bottom) as functions of $\rho$.
Two values of the external potential are considered: $\Dv = -2$ and $\Dv = -1/5$, with each providing a solution on a different branch. 
The stationary densities are indicated by markers.
As one can see, stationary points only appear when the functional is real-valued since the complex part of the potential is excluded by restricting $\Dv$ to real values.
Therefore, unless a complex density is specifically chosen as the starting point for the self-consistent procedure, we can safely limit ourselves to the real part of the functionals without any loss of generality.

%%% FIG 9 %%%

In this context, we define $\vs^{(1)} = \Re[\vs^{(1\pm)}] = 0$ and $\Ts^{(1)} = \Re[\Ts^{(1\pm)}] = 0$ and the real-valued version of Eq.~\eqref{eq:KS1pm} reads
\begin{equation} \label{eq:KS1}
	\vs^{(1)}(\rho) - v_\text{Hxc}^{(m)}(\rho) - \Dv = 0
\end{equation}
where $m = \{ 0, 1\cup, 1\cap, 2 \}$.
The roots of Eq.~\eqref{eq:KS1} are denoted $\rho^{(1,m)}$ and the real-valued KS energy is
\begin{equation} \label{eq:EKS1}
	\EKS^{(1,m)} = \Ts^{(1)}(\rho^{(1,m)}) + E_\text{Hxc}^{(m)}(\rho^{(1,m)}) + \Dv \rho^{(1,m)}
\end{equation}

%%% FIG 10 %%%
\begin{figure}
\includegraphics[width=\linewidth]{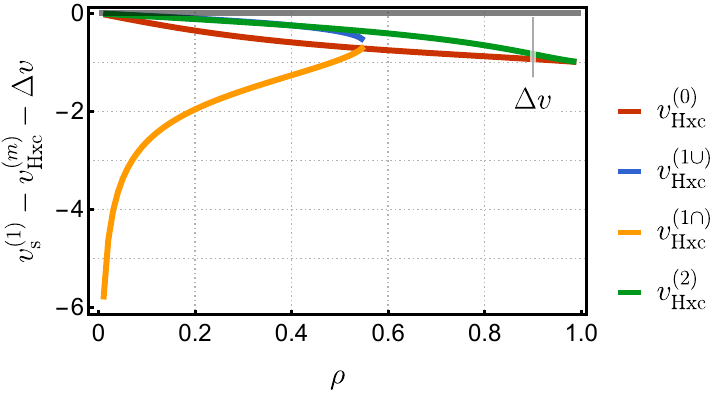}
\caption{$\vs^{(1)} - v_\text{Hxc}^{(m)} - \Dv$ as a function of $\rho$ for $U=1$. 
The horizontal gray line indicates the value of $\Dv$, which is set to zero here. 
Decreasing $\Dv$ would move down this line.
For a given $\Dv$, the stationary densities are given by the intersection of the colored curves and the gray line.}
	 \label{fig:KSpot_Ts1}
\end{figure}

Figure \ref{fig:KSpot_Ts1} illustrates the evolution of $\vs^{(1)} - v_\text{Hxc}^{(m)} - \Dv$ as a function of $\rho$. 
The trends observed here are similar to those in Figs.~\ref{fig:KSpot_Ts0} and \ref{fig:KSpot_Ts2}, and comparable conclusions can be drawn. 
A key distinction, however, is the absence of a stationary solution for $\abs{\Dv} > 1$, except when the concave branch of the first-excited-state functional is employed. 
This underscores the unique character of this functional compared to the others.
As expected, $v_\text{Hxc}^{(1\cup)}$ and $v_\text{Hxc}^{(2)}$ remain remarkably close for small $\Dv$, highlighting their consistency in this regime.

The similarity of $v_\text{Hxc}^{(0)}$ and $v_\text{Hxc}^{(2)}$ can be attributed to the fact that the ground and doubly-excited states describe the same type of electron correlation, namely, two electrons occupying the same orbital.
This correlation is predominantly dynamic in nature and also applies to the first excited state for small $\Dv$ (i.e., the convex branch).
For large $\Dv$, the ground and doubly-excited states can be interpreted as ionic configurations, where both electrons are located on the same site. 
In contrast, in the first excited state at larger $\Dv$ (i.e., the concave branch), the electrons are distributed across different sites, with one electron located on each site. 
This corresponds to a charge-transfer state characterized by covalent configurations.
This distinction highlights the significant difference in the correlation functional between short-range dynamic correlation, where electrons are close to one another, and charge-transfer states, where electrons are distributed across different centers.

%%%%%%%%%%%%%%%%%%%%%%%%
\section{Conclusion}
\label{sec:ccl}
%%%%%%%%%%%%%%%%%%%%%%%%
In this study, we extended the KS-DFT formalism to explore its applicability to excited states, focusing on the asymmetric Hubbard dimer at half-filling. 
While the highest-lying doubly-excited state fits within the KS-DFT framework, the first excited state poses significant challenges due to its lack of non-interacting $v$-representability. 
By employing an analytic continuation of the adiabatic connection, we showed that the density of the first excited state can be generated by a complex-valued external potential in the non-interacting case.
One could regard the density of the first excited state as ``complex-$v$-representable''. 
Our findings might prove useful in systems where the KS $v$-representability condition is violated, such as first- and second-row atoms with partially filled $p$-shells. \cite{TruErhGor-PRA-24}

From a more practical point of view, our results highlight the emergence of spurious stationary solutions in the KS equations when approximate correlation functionals are employed, highlighting the importance of developing more accurate functional approximations for state-specific KS calculations. 
In the case of the Hubbard dimer, this occurs when performing a doubly-excited-state calculation using the ground-state functional, a widely adopted custom among OO-DFT practitioners.
This work provides a deeper understanding of the theoretical and practical limitations of KS-DFT for excited states.

To advance the applicability of OO-DFT, the creation of state-specific functionals is thus essential.\cite{GouPit-PRX-24,GouDalKroPit-arXiv-24}
Such developments would bridge the gap between theoretical insights and practical computations, allowing for more reliable computational tools across chemistry, physics, and material science.

%%%%%%%%%%%%%%%%%%%%%%%%
\section*{Acknowledgements}
%The authors thanks stimulating discussions with Tim Gould, Hugh Burton, and Emmanuel Fromager.
This project has received funding from the European Research Council (ERC) under the European Union's Horizon 2020 research and innovation programme (Grant agreement No.~863481) and under the Marie Skłodowska-Curie grant agreement No.~101104947.
%%%%%%%%%%%%%%%%%%%%%%%%

%%%%%%%%%%%%%%%%%%%%%%%%%%%%%%%%
\section*{Data availability statement}
%%%%%%%%%%%%%%%%%%%%%%%%%%%%%%%%
Data sharing is not applicable to this article as no new data were created or analyzed in this study.

%%%%%%%%%%%%%%%%%%%%%%%%%%%%%%%%
\section*{References}
%%%%%%%%%%%%%%%%%%%%%%%%%%%%%%%%

%%%%%%%%%%%%%%%%%%%%%%%%
\bibliography{KS_Hubbard}
%%%%%%%%%%%%%%%%%%%%%%%%

\end{document}